\documentclass[aps,prl,showpacs,twocolumn,nofootinbib,
superscriptaddress]{revtex4-2}
\usepackage{amsmath,amssymb}
\usepackage{subfigure}

\usepackage{slashed}
\usepackage{amsfonts}
\usepackage{epsf}
\usepackage{rotating}
\usepackage{graphicx}
\usepackage{amsmath}
\usepackage{fancyhdr}
\usepackage{lineno}
\usepackage{braket}
\usepackage{stfloats}
\usepackage[T1]{fontenc} % necessary for "old text encoding" apparently (like 
\usepackage[utf8]{inputenc}
\usepackage{babel}
\usepackage{nicefrac}
\usepackage{graphics}
\usepackage[normalem]{ulem}
\usepackage{pstricks}
\usepackage{color}
\usepackage{multirow}
\usepackage{url}
\usepackage{empheq}
\usepackage[title]{appendix}
\usepackage[compat=1.1.0]{tikz-feynhand}
\usepackage[normalem]{ulem}

\def\sss{\scriptscriptstyle}
\def\nn{\nonumber}
\def\CP{$C\!P$~}

\def\mev{\ensuremath{\mathrm{Me\kern -0.1em V}}}

\allowdisplaybreaks

\begin{document}

\title{Measuring $\mathbf{CP}$ violating phase in beauty baryon decays}

\author{Rahul Sinha}
\email{sinha@imsc.res.in}
\affiliation{The Institute of Mathematical Sciences, Taramani,
		Chennai 600113, India}
\affiliation{Homi Bhabha National Institute, Training School Complex,
		Anushaktinagar, Mumbai  400094, India}

\author{Shibasis Roy} 
\email{shibasis.cmi@gmail.com}
\affiliation{Department of Physics, University of Calcutta, 92 Acharya Prafulla 
Chandra Road, Kolkata 700009, India}

\author{N.~G.~Deshpande}
\email{desh@uoregon.edu}
\affiliation{Institute for Fundamental Science, University of Oregon,
		Eugene, Oregon 94703, USA}

	\date{\today}

\begin{abstract}
One of the outstanding problems in physics is to explain the baryon-anti-baryon
asymmetry observed in nature. According to the well-known Sakharov criterion for
explaining the observed asymmetry, it is essential that \CP violation exist. 
Even though \CP violation has been observed in meson decays and is an integral
part of the standard model (SM), measurements in meson decays indicate that, \CP
violation in  the SM is insufficient to explain the observed baryon-anti-baryon
asymmetry. SM predicts the existence of yet to be observed \CP
violation in baryon decays.  A critical test of the SM requires that \CP 
violation be measured in baryon decays as well, in order to verify that it 
agrees with the measurement using meson decays. In this letter  we propose a 
new method to measure \CP violating phase in $b$-baryons, using interference 
arising implicitly due to Bose symmetry considerations of the decaying 
amplitudes.
\end{abstract}

\maketitle

The proposal to observe \CP violation in $\Sigma$-baryon
decays~\cite{Okubo:1958zza} was first made by Okubo in 1957, much before the
first observation of \CP violation  in 1964 by Fitch and Cronin using neutral
$K$-mesons. \CP violation has since been observed in $B$-meson and $D$-mesons as
well. \CP violating phases have been measured accurately in $B-$meson decays and
the measurements are found consistent within the standard model (SM) and in
agreement the Kobayashi-Maskawa hypothesis.

One of the well-known Sakharov conditions~\cite{Sakharov:1967dj,Sakharov:1979xt}
to explain baryon-anti-baryon asymmetry in the universe, is the existence of \CP
violation. The observation of \CP-violation in baryon decays is expected within
the SM and thus, of great interest. It would provide a direct probe of
matter-antimatter asymmetry, arising from \CP violating sector of the SM. This 
is
essential since measurements in meson decays indicate that, \CP violation in 
the SM is by itself insufficient to explain the observed baryon-anti-baryon
asymmetry~\cite{Davidson:2008bu}.  However, \CP violation has only been observed
in mesons decays and is yet to be convincingly demonstrated
\cite{Aaltonen:2011qt,Aaltonen:2014vra, Aaij:2016cla, Aaij:2018lsx,
Aaij:2018tlk, Aaij:2019rkf,Aaij:2019mmy,Aaij:2019pqz,Cerri:2018ypt,LHCb:2021enr}
in baryon decays.  It is well-known that the largest weak phases within the SM
appear in $b\to u$ and $t\to d$ transitions. It is thus obvious that it would be
easiest to observe \CP violation in $b$-baryons and one must explore the weak
decays of the anti-triplet ($\bar{3}$) beauty baryons.

While  there is no reason to expect a disagreement between \CP violating phases
measured using baryons and that using mesons, within the SM, it nevertheless
remains imperative, that a clean measurement of the weak phase also be performed
in baryons decays. A comparison of the measurements of \CP violation in meson
and baryon decays would provide a critical probe of physics beyond the SM.
Unfortunately, measurement of \CP violation in baryon decays is not straight
forward. It is well-known that for \CP violation to be observed the amplitude
must have two contributions with both having different strong phases and weak
phases. However, in order to {\em measure \CP violating phase} using the
decay(s) of any particle(s) one must satisfy certain further conditions. One of
the critical requirements to measure any phase is that amplitudes interfere. In
the case of neutral $B$-mesons mixing between particle and antiparticle allows
for two distinct amplitudes to interfere, with one amplitude corresponding to
direct decay and the other to decay via mixing. This results in the well-known
time dependent \CP violation. However, baryon number conservation forbids
oscillations between baryons and anti-baryons disallowing such time dependent
mixing and consequent interference of two amplitudes. It is also essential that
at least one of the decay amplitudes be re-parametrization
invariant~\cite{London:1999iv, Imbeault:2006nx} such that  the \CP violating
phase to be measured can be uniquely defined~\cite{foot:reparameterization}.
Inability to satisfy this condition results in fewer independent observables
than necessary to measure the weak phase. The methods to measure the phase
$\beta$ and $\alpha$ without approximations, in $B$-mesons satisfy all these
conditions.

Interestingly, the phase $\gamma$ can be measured, without the need for meson
mixing. Interference of two decay amplitudes arises from decay to the same final
state via intermediate $D^0$ and $\bar{D}^0$ decays. The only
proposal~\cite{Giri:2001ju} to measure weak phase using baryons is a
straightforward extension of this method used in mesons, to achieve interference
of two amplitudes. A proposal to measure \CP violating weak phase  in baryon
decays that is not a straightforward extension of the approach used in meson
decays and one that  relies on processes specific to baryons is still missing.

In this letter we propose a new approach to measure \CP violating weak phase in
baryon decays using interference arising  due to Bose symmetry considerations of
the decaying amplitudes. We start by exploring the weak decays of the
anti-triplet $\bar{3}$ beauty baryons which have been studied earlier in
Ref.~\cite{Roy:2019cky,Roy:2020nyx} and where large \CP violation may be
expected.  As mentioned earlier for weak-phases to be measurable, the additional
criterion of re-parametrization invariance must also hold. This
re-parametrization condition is severely restrictive for the weak decay of
$\bar{3}~b$-baryon and we find that even when the interference criterion due to
Bose symmetry holds, measurement of the weak phase is not possible in most cases
since re-parametrization invariance of the amplitude does not hold. Considering
all the weak decay modes of the $\bar{3}$ $b$-baryons we find one mode namely,
$\Xi_b\to \Sigma^\prime(1385)\pi$, where it is possible to measure the \CP 
violating weak phase.

\begin{figure}[h]
%\centering
\includegraphics[scale=0.75]{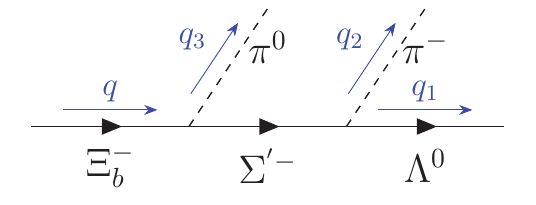}
\includegraphics[scale=0.75]{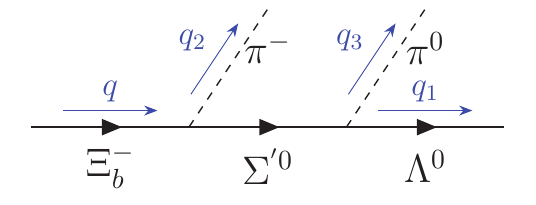}
\caption{Feynman diagrams contributing to the quasi two body decay  
$\Xi_b^-(q)\to \Lambda^0(q_1)\pi^-(q_2)\pi^0(q_3)$ via the 
$J^P=\tfrac{3}{2}^+$,
$\Sigma^\prime(1385)$ baryon resonance.}
\label{Fig:diagram1}
\end{figure}

We show in detail how the weak phase can be measured using the decay $\Xi_b\to
\Sigma^\prime\pi$, where $\Sigma^\prime$ is the $J^P=\tfrac{3}{2}^+$ decuplet 
baryon resonance
corresponding to $\Sigma^\prime(1385)$. We consider the decays
$\Xi_b^-\to\Sigma^{\prime 0}(1385)\pi^-$ and $\Xi_b^-\to\Sigma^{\prime -}(1385) 
\pi^0$
depicted in Fig.~\ref{Fig:diagram1}. Note that $\Xi_b^-$ is an isospin $1/2$
state of $bsd$ quarks.
% We will refer to $\Sigma^\prime(1385)$  simply as $\Sigma$ henceforth. 
The subsequent strong-decays of the $\Sigma^{\prime 0}\to \Lambda^0\pi^0$
and $\Sigma^{\prime -}\to \Lambda^0\pi^-$ result in the identical final state
$\Xi_b^-\to \Lambda^0\pi^-\pi^0$ for both the decay modes. The
$\Lambda^0\pi^-\pi^0$ Dalitz plot would depict both these decays and have
special properties under exchange of the two-pions which are identical 
bosons under isospin symmetry. The interference between
these two modes would arise implicitly from Bose symmetry correlations even
though the two decay modes do not effectively interfere on the Dalitz plot. The
$\pi^-\pi^0$ can either be in a $\left|1,-1\right>$ or $\left|2,-1\right>$
isospin state. Since, the two $\pi$-bosons are identical under isospin their
total wave-function must be symmetric. This necessitates that the two pions in
the odd isospin state $\left|1,-1\right>$ must be anti-symmetric under spatial
exchange, whereas, the two pions of even isospin state $\left|2,-1\right>$ must
be symmetric under spatial exchange.  The $\Lambda^0$ is an isospin
$\left|0,0\right>$ state, hence, the combined isospin of the final
$\Lambda^0\pi^-\pi^0$ state must be identical to the isospin of the two pion
state, which  can either be a $\left|2,-1\right>$ or $\left|1,-1\right>$.
Isolating the $\pi^-\pi^0$ symmetric state which has isospin
$\left|2,-1\right>$, would thus be equivalent to isolating the $\Delta I=3/2$
changing contribution to the decay, which cannot arise from the penguin diagram
and must have a pure $b\to u$ weak phase.

\begin{figure}[h]
\centering 
\includegraphics[scale=0.8]{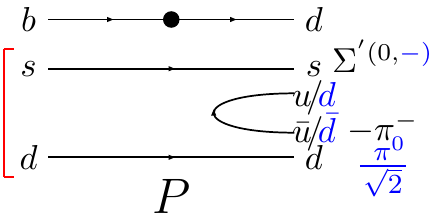}
\includegraphics[scale=0.8]{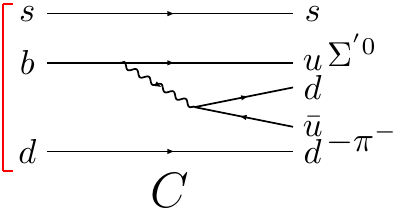} 
\caption{Topological diagrams
contributing to decay $\Xi_b^-\to \Sigma^\prime \pi$ decays. The blob on the 
$b\to d$
transition in the diagram on the left corresponds to the $b\to d$ penguin. The
red line straddling between the $s$ and $d$ quarks of the initial $\Xi_b$ baryon
indicates anti-symmetry of quarks in the initial  $\bar{3}$ state. The 
symmetric wave-function of the $J^P=\tfrac{3}{2}$ nucleon $\Sigma^\prime$ 
allows only 
the above topologies. } 
\label{Fig:Topologies}
\end{figure}

\begin{figure}[b]
%  \centering
\includegraphics[scale=0.4]{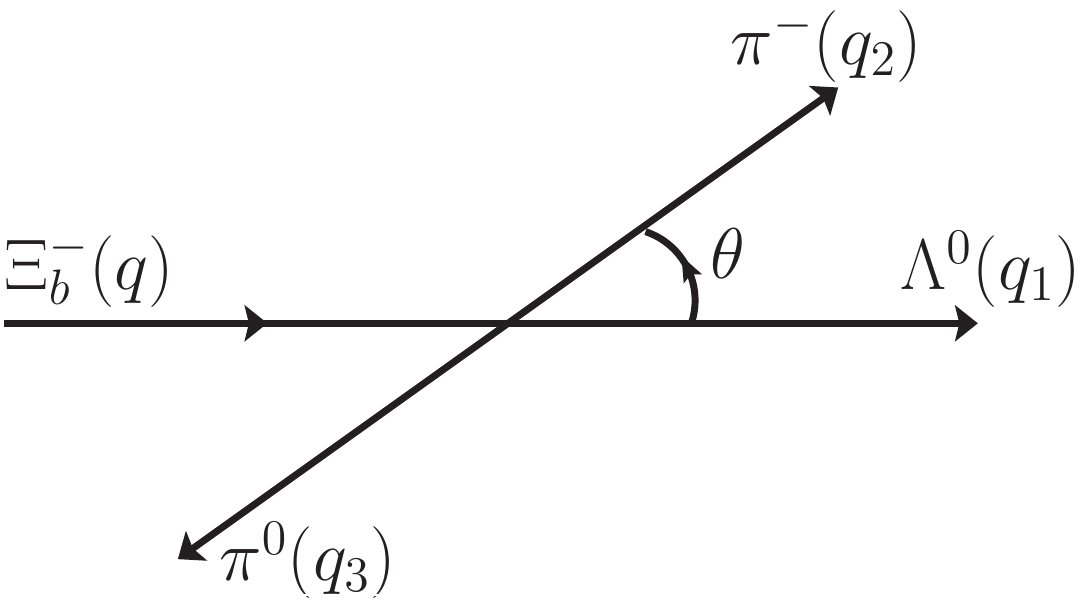}
\caption{Jackson frame in which $\Xi_b^-(q)$ decays to three bodies 
$\Lambda^0(q_1)$, $\pi^-(q_2)$ and $\pi^0(q_3)$~.}
\label{Fig:Jackson}
\end{figure}

The  topological diagram contributions to the decays $\Xi_b^-\to\Sigma^{\prime
0}(1385)\pi^-$ and $\Xi_b^-\to\Sigma^{\prime -}(1385) \pi^0$ are depicted in
Fig.~\ref{Fig:Topologies}. While the decay $\Xi_b^-\to\Sigma^{\prime
0}(1385)\pi^-$  gets contributions from both the tree-level and penguin diagrams
with the corresponding amplitudes denoted as $C$  and  $P$ respectively, only
the $P$ amplitude contributes to the decay  $\Xi_b^-\to\Sigma^{\prime -}(1385)
\pi^0$. This ensures that the weak phase cannot be reparametrizated and plays a
crucial role in the measurement of the \CP violating weak phase. The \CP
violating weak phase measured using $b$-baryons with in our approach  is free
from any hadronic uncertainty and relies only on reliable theoretical inputs
such as isospin and vanishingly small electro-weak penguin contributions in
$\Delta S=0$ $b\to d$ decays.

We consider the decay in the Gottfried-Jackson frame with $\Xi_b$ moving in 
the $+\hat{z}$ axis, with the two pions going back to back with the 
$\pi^-(q_2)$ at an angle $\theta$ to the $\Lambda^0(q_1)$. In this 
frame (see 
Fig.~\ref{Fig:Jackson}) $\vec{q_2}+\vec{q_3}=0$. We define 
$s\equiv(q_2+q_3)^2=(q-q_1)^2$, 
$t\equiv(q_1+q_3)^2=(q-q_2)^2$ and $u\equiv(q_1+q_2)^2=(q-q_3)^2$. $t$ and $u$ 
can be written as 
\begin{align}
t&=x+y\cos\theta\label{eq:t}\\
u&=x-y\cos\theta
\label{eq:u}
\end{align}
where 
\begin{gather}
x=\frac{M^2+m_\Lambda^2+2 m_\pi^2-s}{2}\label{eq:x}\\
y=\frac{\sqrt{s-4 
m_\pi^2}}{2\sqrt{s}}\lambda^{\nicefrac{1}{2}}(M^2,m_\Lambda^2,s)\label{eq:y},
\end{gather}
and $\lambda(M^2,m_\Lambda^2,s)=(M^4+m_\Lambda^4+s^2-2 M^2 m_\Lambda^2-2 M^2 
s-2 m_\Lambda^2 s)$. $M$ is the mass of the $\Xi_b$ baryon, $m_\Lambda$ is the 
mass of the $\Lambda^0$ and $m_\pi$ is the mass of the pions.

The matrix element for the weak-decay $\Xi_b^-\to 
\Sigma^{\prime(-,0)}\pi^{(0,-)}$
is given by
\begin{equation}
{\cal M}(\Xi_b\to\Sigma^\prime\pi)=-i q^\pi_\mu 
\bar{u}^\mu_{\sss\Sigma^\prime}(a+b\gamma_5)u_{\sss\Xi_b},
\end{equation}
where, $u^\mu_{\sss\Sigma^\prime}$ is the Rarita-Schwinger spinor for the
spin-$\nicefrac{3}{2}$ decuplet and hence a Lorentz index, $q^\pi_\mu$ is the
momentum of the $\pi$, and $u_{\sss\Xi_b}$ is the spinor of the $\Xi_b$. The
two coefficients $a$ and $b$ depend on the CKM elements and flavor structure
specific to the decay mode. It may be noted that $a$ and $b$ are related to the
$p$- and $d$-wave decay amplitudes respectively. The 
$\Sigma^\prime$
baryon subsequently decays via strong-interaction to a $\Lambda^0$-baryon and a 
pion with
the matrix element for the decay being given by,
\begin{equation}
{\cal M}(\Sigma^\prime\to \Lambda^0\pi)=i g_{\sss \Sigma^\prime \Lambda\pi}  
q^\pi_\mu 
\bar{u}_\Lambda u^\mu_\Sigma,
\end{equation}
where, $g_{\sss \Sigma^\prime \Lambda\pi}$ is the invariant coupling for the 
decay. The
decay for the conjugate modes, 
$\overline{\Xi}_b^+\to\overline{\Sigma}^{\prime(+,0)}\pi^{(0,+)}$ can 
be 
described analogously except that
it would be described by new coefficients $\overline{a}$ and $\overline{b}$ that
are related to $a$ and $b$ by reversing the respective weak phases. The 
propagator of the $\Sigma^\prime(1385)$ baryon  has the form \cite{Choi:1989yf}
\begin{align}
\Pi^{\mu\nu}(k)&=-\frac{(\slashed{k}+m)}{(k^2-m^2+im \Gamma)}
\Big(g^{\mu\nu} -\frac{2}{3} \frac{k^\mu k^\nu}{m^2}  \nn \\ &\qquad 
-\frac{1}{3} \gamma^\mu \gamma^\nu +\frac{1}{3m}
(k^\nu\gamma^\mu-k^\mu\gamma^\nu)\Big),
\end{align}
corresponding to that of a spin-$\nicefrac{3}{2}$ fermion with the four-momentum
$k$, mass $m$ and width $\Gamma$.

The matrix element for the two-step decay $\Xi_b^-(q)\to
\Sigma^{\prime -}[\to \Lambda^0(q_1)\pi^-(q_2)] \pi^0(q_3)$ is given by
\begin{multline}
\label{eq:Mu}
{\cal M}_u={\cal M}\Big(\Xi_b^-(q)\to \Sigma^{\prime-}\big[\to
\Lambda^0(q_1)\pi^-(q_2)\big] \pi^0(q_3)\Big)=\\
g_{\sss \Sigma^\prime \Lambda\pi} 
\bar{u}(q_1)(a^{\sss-}+b^{\sss-}\gamma_5) 
\Pi^{\mu\nu}(q_{12}) 
u(q)\;q_3^\mu q_2^\nu,
\end{multline}
where $q_{ij}=q_i+q_j$ and $m$ is the mass of $\Sigma^\prime(1385)$ resonance. 
Similarly, the two-step decay $\Xi_b^-(q)\to\Sigma^{\prime0}[\to \Lambda^0(q_1)
\pi^0(q_3)] \pi^-(q_2)$ is given by
\begin{multline}
\label{eq:Mt}
{\cal M}_t={\cal M}\Big(\Xi_b^-(q)\to \Sigma^{\prime 0}\big[\to
\Lambda^0(q_1)\pi^0(q_3)\big] \pi^-(q_2)\Big)= \\
g_{\sss \Sigma^\prime \Lambda\pi}\bar{u}(q_1)(a^{\sss0}+b^{\sss0}\gamma_5) 
\Pi^{\mu\nu}(q_{13})
u(q)\;q_2^\mu q_3^\nu.
\end{multline}
The matrix elements ${\cal M}_u$ and ${\cal M}_t$ are related by isospin. The
sum of the matrix element of the two contributing modes must be completely Bose 
symmetric
under the exchange of the two pions and may be written in an explicitly 
symmetric form as
%\begin{multline}
%{\cal M}\Big(\Xi_b^-\to \Sigma\big[ 
%\Lambda\pi\big] \pi\Big)= g_{\sss \Sigma \Lambda\pi} \bar{u}(q_1)\\
%\Big[(A_{e}+B_{e}\gamma_5)
%\big(\Pi^{\nu\mu}(q_{12}) +\Pi^{\mu\nu}(q_{13})\big) \\ 
%+(A_{o}+B_{o}\gamma_5)
%\big(\Pi^{\nu\mu}(q_{12})-\Pi^{\mu\nu}(q_{13})\big)\Big]u(q) q_2^\mu 
%q_3^\nu \label{eq:Bose}
%\end{multline}
\begin{multline}
{\cal M}\Big(\Xi_b^-\to \Sigma^\prime\big[ 
\Lambda\pi\big] \pi\Big)= g_{\sss \Sigma^\prime \Lambda\pi} \bar{u}(q_1)
\Big[(A_{e}+B_{e}\gamma_5)\\
\qquad\qquad\big(\Pi^{\nu\mu}(q_{12}) +\Pi^{\mu\nu}(q_{13})\big)
+(A_{o}+B_{o}\gamma_5) \\ 
\big(\Pi^{\nu\mu}(q_{12})-\Pi^{\mu\nu}(q_{13})\big)\Big]u(q) q_2^\mu 
q_3^\nu \label{eq:Bose}
\end{multline}
where $A_e$, $B_e$ and $A_o$, $B_o$ are the  even and odd parts of the 
amplitude under the 
exchange of the two pions and are given by
\begin{align}
A_{e,o}=&(a^{\sss-}\pm a^{\sss0})/2\nn \\
B_{e,o}=&(b^{\sss-}\pm b^{\sss0})/2.
\end{align}

\begin{figure}[b]
\includegraphics[scale=0.215]{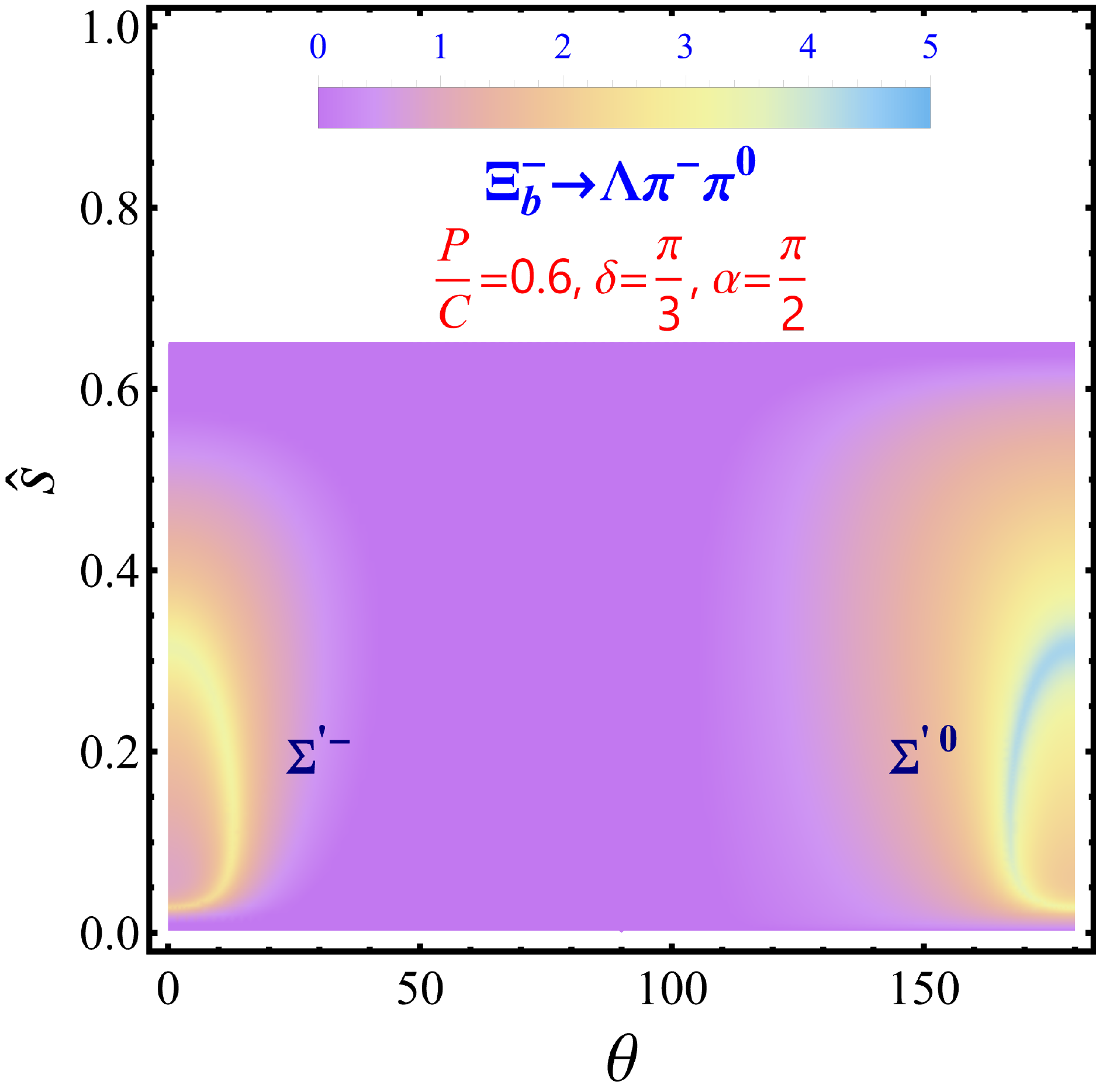} 
\includegraphics[scale=0.215]{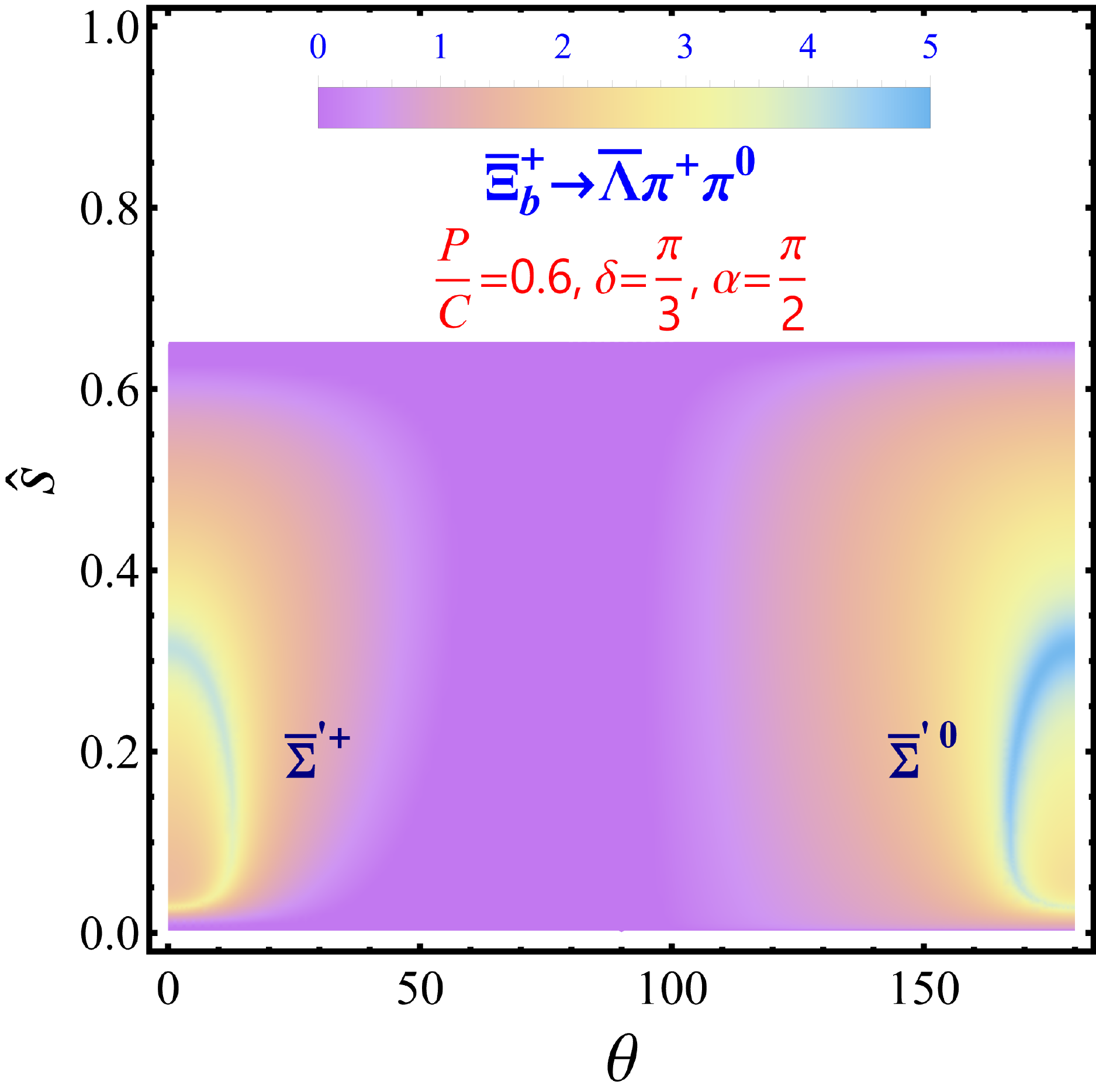} 
\caption{ %
Sample correlation-plots for the $p$-wave contributions for the
specified choice of parameters. Plotted are the logarithmic value of the
$p$-wave rates with an arbitrary scale that depends on $C$. Explicit
dependence of the some of the parameters on $p$-wave is suppressed for
simplicity. The contributing intermediate $\Sigma^\prime$ resonances are 
depicted on the plots.
%The $\Sigma^-$ and $\Sigma^0$ resonances are evident in the plot on the left 
%and $\bar{\Sigma}^+$ and $\bar{\Sigma}^0$ resoances on the right plot 
%respectively. 
The differences between the two plots arise only due \CP 
violation effects. Similar plots would arise from the $d$-wave
contributions. } %
\label{Fig:Dalitz-1}
\end{figure}

The coefficients $a^{\sss0}$, $a^{\sss-}$, $b^{\sss0}$ and $b^{\sss-}$ are
related  (see Fig.~\ref{Fig:Topologies}) to the topological amplitudes
\cite{Roy:2019cky,Roy:2020nyx},  which after a phase rotation of $e^{i\gamma}$ 
may be 
written
as follows
\begin{align}
\label{eq:P&C}
a^{\sss0}=& 
-\tfrac{1}{2\sqrt{3}}\big(C_{p}-P_{p}e^{-i\alpha}\big),\nn\\
a^{\sss-}=&-\tfrac{1}{2\sqrt{3}} P_{p}e^{-i\alpha}\nn\\
b^{\sss0}=&
-\tfrac{1}{2\sqrt{3}}\big(C_{d}-P_{d}e^{-i\alpha}\big),\nn\\
b^{\sss-}=& -\tfrac{1}{2\sqrt{3}} P_{d}e^{-i\alpha},
\end{align}
where $C_{p}$, $P_{p}$ are related to the topological
amplitudes contributing to the ${p}$-wave with analogous definitions
for the ${d}$-wave. Here the $p$-wave and $d$-wave refer to waves in the 
intermediate $\Sigma^\prime \pi$ state. Note that the phase rotation of 
$e^{i\gamma}$ does
not alter any physical observable. Hence,
\begin{align}
\label{eq:xyz}
A_{e}=& -\tfrac{1}{4\sqrt{3}} C_{p}
	\equiv   x_{p} \nn \\
A_{o}=& \tfrac{1}{4\sqrt{3}} C_{p}
-\tfrac{1}{2\sqrt{3}} P_{p}e^{-i\alpha}	\equiv -x_{\sss p}
+z_{ p} e^{-i\alpha} e^{i\delta_{ p}} \nn \\
B_{e}=& -\tfrac{1}{4\sqrt{3}} C_{d}
	\equiv  x_{d}\nn \\
B_{o}=& \tfrac{1}{4\sqrt{3}} C_{d}
-\tfrac{1}{2\sqrt{3}} P_{d}e^{-i\alpha}	
	\equiv -x_{d}+z_{d} 
	e^{-i\alpha} e^{i\delta_d},
\end{align}
where, $\delta_{(p,d)}$, is the strong phase difference between the penguin and 
the 
tree contribution for the  ${p}$- and  ${d}$-waves respectively.

The decay rate for the 3-body final state $\Lambda^0\pi^-\pi^0$ will depend on
the Mandelstam variables $t$ and $u$. This can easily be cast in terms of $s$
and $\cos\theta$ using Eqs.~\eqref{eq:t},~\eqref{eq:u},~\eqref{eq:x} and
\eqref{eq:y}. It is obvious to conclude from Fig.~\ref{Fig:Jackson} that under
the exchange of the two pions, $\theta \leftrightarrow \pi-\theta$. The odd
(even) part of the amplitude under the exchange of two pions must therefore be
proportional to odd (even) powers of $\cos\theta$.  Since, Bose correlations are
clearly depicted in plots involving $\hat{s}=s/M^2$ and $\theta$, we refer to
such plots as correlation-plots. In Fig~\ref{Fig:Dalitz-1}, we show sample
correlation-plots for $P_p/C_p=0.6$, $\delta_p=\pi/3$
and $\alpha=\pi/2$. Plotted are the logarithmic values of the $p$-wave
rates with an arbitrary scale that depends on the choice of
$P_p/C_p$ and the $p$-wave rate for the decay
$\Xi_b^- \to \Sigma^{\prime-}\pi^0$.  In our calculations we have used 
$M=5797$~\mev,
$m=1385$~\mev, $\Gamma=38$~\mev, $m_\Lambda=1115$~\mev and $m_\pi=135$~\mev. The
effects of \CP violation can be observed as differences between the two plots
corresponding to mode and conjugate mode. The effects of Bose correlation are
also obvious. The decay $\Xi_b^-\to \Sigma^{\prime-}\pi^0$ arises due to pure 
penguin (See
Eqns.~\eqref{eq:Mu} and \eqref{eq:P&C}). If there are no  Bose correlation
effects contributing to the decays, the decays of $\Xi_b^-\to 
\Sigma^{\prime-}\pi^0$ and
$\overline{\Xi}_b^+\to \overline{\Sigma}^{\prime+}\pi^0$ seen on the left of 
the two 
plots in
Fig.~\ref{Fig:Dalitz-1} would look identical. However, the Dalitz plots show a
marked difference between  $\Xi_b^-\to \Sigma^{\prime-}\pi^0$ and 
$\overline{\Xi}_b^+\to
\overline{\Sigma}^{\prime +}\pi^0$ on their respective Dalitz plot, {\em 
providing a 
smoking
gun evidence of large Bose correlation effects}. This observation is fundamental
to our new approach to measure \CP violation. We will show next that {\em the
observed distributions along these narrow resonances provide enough information
to extract all theoretical parameters from experimental data}, including the
${p}$- and ${d}$-wave contributions respectively.

The numerator of  the decay rate $N_\Gamma$ for both the mode and conjugate 
mode are worked out to have the 
complicated form:
\begin{equation}
N_\Gamma=\sum_{n=0}^{4}c_{n}(\hat{s}) 
\cos2n\theta+\sum_{n=0}^{3}d_{n}(\hat{s})
\cos(2n+1)\theta,
\label{eq:Gamma}
\end{equation}
where, all masses and momenta are normalized to $M$, the
$\Xi_b$ mass for simplicity, and
\begin{align}
\label{eq:cn}
&c_{n}(s)=\big(f_{n}^{\sss(1)}(\hat{s})|A_{e}|^2+ 
f_{n}^{\sss(2)}(\hat{s})|B_{e}|^2 \nn\\
	&\qquad\quad+f_{n}^{\sss(3)}(\hat{s})|A_{o}|^2 + f_{n}^{\sss(4)}(\hat{s}) 
	|B_{o}|^2 \big) 
	\\ 
&d_{n}(\hat{s})=\big(g_{n}^{\sss(1)}(\hat{s})\text{Re}(A_eA_o^*)+
	g_{n}^{\sss(2)}(\hat{s})\text{Re}(B_eB_o^*)\nn \\ 
	&\qquad\quad+g_{n}^{\sss(3)}(\hat{s})\text{Im}(A_eA_o^*)
	+g_{n}^{\sss(4)}(\hat{s})\text{Im}(B_eB_o^*)\big).
\label{eq:dn}
\end{align}
The coefficients $f_{n}^{(i)}$ and $g_{n}^{(i)}$ are functions of $\hat{s}$ and
are easily computed in terms of kinematic factors. For a given choice of
$\hat{s}$, $f_{n}^{(i)}$ and $g_{n}^{(i)}$ are just numbers. We henceforth drop
explicit dependence on $\hat{s}$, since our solutions are valid for all
$\hat{s}$. The numerator of the decay rate given in Eq.~\eqref{eq:Gamma} can be
fitted as a function of $\theta$ to obtain coefficients $c_n$ and $d_n$. Having
obtained $c_0$, $c_1$, $c_2$ and $c_3$ it is obvious that $|A_{e}|^2$,
$|A_{o}|^2$, $|B_{e}|^2$ and $|B_{o}|^2$ can easily be obtained. Similarly,
$\text{Re}(A_eA_o^*)$, $\text{Re}(B_eB_o^*)$, $\text{Im}(A_eA_o^*)$ and
$\text{Im}(B_eB_o^*)$ can be solved using $d_0$, $d_1$, $d_2$ and $d_3$. In
order to solve for the amplitudes and their interference, one must have a
minimum of 8-bins in $\theta$ and $\hat{s}$ involving both the resonances
contributing to the process. Experimental procedure used in an actual analysis
can be more refined. An identical analysis of the decay rate for the conjugate
process would enable us to solve for $|\overline{A}_e|^2$, $|\overline{A}_o|^2$,
$|\overline{B}_e|^2$, $|\overline{B}_o|^2$ and
$\text{Re}(\overline{A}_e\overline{A}_o^*)$,
$\text{Re}(\overline{B}_e\overline{B}_o^*)$,
$\text{Im}(\overline{A}_e\overline{A}_o^*)$ and
$\text{Im}(\overline{B}_e\overline{B}_o^*)$. Having solved the values of these
amplitudes and their interference, our aim is to solve for the weak phase
$\alpha$ and the amplitudes $x_{{p,d}}$ and $z_{{p,d}}$
defined in Eq.~\eqref{eq:xyz} and the strong phases $\delta_{{p,d}}$.
In order to obtain the solutions for the $p$-wave parameters we define
new intermediate observables $r_i$:
\begin{align}
r_0=& |A_e|^2=|\overline{A}_e|^2=x_p^2\nn\\
r_1=& |A_o|^2+|\overline{A}_o|^2
	=2 x_p^2+2 z_p^2-4 x_p z_p 
	\cos\delta_p \cos\alpha\nn\\
r_2=& |A_o|^2-|\overline{A}_o|^2
	=-4\, x_p z_p \sin\delta_p \sin\alpha \nn\\
r_3=&\text{Re}(A_eA_o^*-\overline{A}_e\overline{A}_o^*
	)=2 x_p z_p \sin\delta_p\sin\alpha\nn\\
r_4=&\text{Im}(A_eA_o^*-\overline{A}_e\overline{A}_o^*)=2 x_p 
z_p \cos\delta_p\sin\alpha \nn\\
r_5=&\text{Re}(A_eA_o^*+\overline{A}_e\overline{A}_o^*) 
	=-2 x_p^2  + 2 x_p z_p 
	\cos\delta_p\cos\alpha \nn\\
r_6=&\text{Im}(A_eA_o^*+\overline{A}_e\overline{A}_o^*) 
	=-2 x_p z_p \sin\delta_p\cos\alpha .
\end{align}
We find the solutions in terms of $r_i$ to be,
\begin{align}
 x_p^2&= r_0  \nn\\
\tan\delta_p&=r_3/r_4 \nn\\
\tan\alpha&=-r_3/r_6= r_2/(2 r_6)\nn\\
z_p^2&=\frac{(r_3^2 + r_4^2) (r_3^2 + r_6^2)}{4 r_0 r_3^2}
\end{align}
Since, all the $r_i$ are not independent, we find the relations:
\begin{align}
r_3^2 + r_4^2 + r_5^2 + r_6^2& = 2 r_0  r_1\nn\\
r_3 r_5 + r_4 r_6 &= r_0 r_2\\
r_2=-2 r_3
\end{align} 
The solutions for the $d$-wave parameters can be
obtained similarly. It is noted that both the $p$- and $d$-
wave contributions must result in the measurement of the same weak phase
$\alpha$. We hence have two independent measurements of $\alpha$ corresponding
to the two partial waves.

The observed Dalitz plot must have a more complicated structure with several
resonances, and one may wonder if our approach to measure the weak phase would
be possible. The contribution from heavier $\Sigma^\prime$ states have a 
similar decay
dynamics with the same weak phase $\alpha$, but the relevant amplitudes and
strong phases would differ. While these resonances are not a cause for concern
suitable binning cuts would easily remove their contributions without losing
relevant signal sample. Only the interference with the decay mode $\Xi_b^-\to
\Lambda^0\rho^-\to \Lambda^0\pi^-\pi^0$, and modes involving heavier $\rho$-like
resonances, instead of $\rho$, need a closer look. It is interesting to note
that $\Xi_b^-\to \Lambda^0\rho^-$  has a different $\hat{s}$ dependence in the
overlap region and contributes  only to the odd part of the amplitude. Once data
is available in several more $\hat{s}$ bins, such interference effects can
easily be isolated using the Dalitz distribution ~\cite{footnote}. One does not
need to impose cuts to remove the $\Xi_b^-\to \Lambda^0\rho^-$ contributions.
 
We have demonstrated that  Bose correlations arise from two intermediate decays
$\Xi_b^-\to \Sigma^{\prime0}\pi^-$ and $\Xi_b^-\to \Sigma^{\prime-}\pi^0$ 
contributing to the
final state $\Xi_b^-\to \Lambda^0\pi^-\pi^0$. Similar Bose correlations arise in
the decay to conjugate final state $\overline{\Xi}_b^+\to 
\overline{\Lambda}^0\pi^+
\pi^0$. We expliticly show that the weak phase $\alpha$ can be measured using
both even and odd contributions to the amplitudes under pion exchange and 
comparing the mode and conjugate-mode correlation-plots. Naively,
one may be tempted to incorrectly assume that Bose correlation effects are too 
small to be
observed.  However, we have shown numerically and explicitly that such
correlations are sizable providing an unequivocal approach to measuring \CP
violation in baryon decays. {\em The significant difference between the  pure 
penguin
decays of $\Xi_b^-\to \Sigma^{\prime-}\pi^0$ and $\overline{\Xi}_b^+\to 
\overline{\Sigma}^{\prime+}\pi^0$ 
on their respective Dalitz plot is a smoking gun evidence of large Bose
correlation effects}. Our new approach to measure \CP violation critically
depends on such Bose correlations. Interestingly, inexplicably large \CP
violation effects are observed in the Dalitz plot of three body $B$-meson
decays; we conjecture that these arise due to Bose correlations on the Dalitz
plot.

We thank Sheldon Stone and Marcin Chrz\k{a}szcz for reading the manuscript and
providing valuable suggestions. The work of SR is supported by the grant
CRG/2019/000362, Science and Engineering Research Board, Government of India.

{\it We dedicate this letter to the memory of Sheldon Stone. He went through 
our draft even while he was in hospital and provided useful comments; such was 
his love for physics. His forthright attitude and recognition of good physics 
won support from many physicists. His presence will forever be missed.
}


\begin{thebibliography}{99}
\bibitem{foot:reparameterization}
Consider the amplitude $A=a+b e^{i\phi}$ which can be re-expressed as $A= 
a+a_0+b_0 
e^{i\phi_0}$ (where $a$, $b$, $a_0$ and $b_0$ are complex amplitudes on account 
of the strong phase). Note that the amplitudes for the mode and conjugate mode 
remain 
invariant 
even when the weak phase has been reparametrized $\phi\to\phi_0$ by also 
modifying 
$a\to 
a+a_0$ and $b\to b_0$. Since $a$, $b$ and $\phi$ are all unknown the phase 
$\phi$ cannot be uniquely defined. The  amplitude for $\Xi^-\to 
\Sigma^{\prime-}\pi^0$ does not allow reparametrization in the defination of 
the 
weak phase (see Eq.~\eqref{eq:P&C}) enabling a solution for the weak phase.
\bibitem{footnote} It is well-known that contributions from two interfering
resonances can be separated using the distribution of events on the Dalitz plot.
A detailed discussion of such an analysis will be presented elsewhere.
\end{thebibliography}
\end{document}